\documentclass[pre,aps,showpacs,preprint]{revtex4-1}

\newcommand{\beq}{\begin{equation}}
\newcommand{\eeq}{\end{equation}}
\newcommand{\bqn}{\begin{eqnarray}}
\newcommand{\eqn}{\end{eqnarray}}
\newcommand{\non}{\nonumber}
\usepackage{graphicx}
\usepackage{epstopdf}
\newcommand{\myscalebox}[1]{\scalebox{0.60}[0.70]{#1}}

\begin{document}

\title{  A simple method to construct eigenset of single-active-electron atom in momentum space with applications  to solve time-dependent Schr\"{o}dinger equation}

\author{Shih-Da Jheng and Tsin-Fu Jiang}
\email{tfjiang@nctu.edu.tw}
\affiliation{Institute of Physics, National Chiao Tung University, Hsinchu 30010, Taiwan}
\date{\today}

\begin{abstract}
 We present a highly accurate method for solving single-active-electron
 (SAE) atomic eigenset in momentum space. The trouble of Coulomb kernel singularity is bypassed with numerical quadrature, which is simple but effective.
  The complicated Lande regularization method is no longer necessary. The data of accuracy for some low-lying states of the hydrogen and SAE helium atom were tabulated.   Two examples of using the generated eigenset to solve the hydrogen atom under strong-field  laser pulses were shown. The momentum and the coordinate representation are  complementary to each other in quantum mechanics. The simple method to generate eigenstates and the localized behavior of wave functions in momentum space would be useful in the study of quantum mechanical problems involving  continuous states.
\end{abstract}

\maketitle

\section{Introduction}
 The continuous eigenstates of a Schr\"{o}dinger equation usually spread over a large volume in coordinate space ({\em R}-space).  From the complementary principle of quantum mechanics, the corresponding momentum space ({\em P}-space) volume will be finite and small for continuous states. Hence, solving quantum mechanical problems in momentum representation have the numerical advantage, especially for
 wide-spreading states in {\em R}-space \cite{chu,jcp}.
 Although we can easily calculate low-lying bound states with modest grids and
 finite range in {\em R}-space, but for problems involved continuous states,
 special efforts are required. Some sophisticated methods can be found in \cite{r-space,toru,tong}.
 In the early years, Podolanski and Pauling \cite{pauling} calculated the Fourier transform of the hydrogen {\em R}-space bound states into {\em P}-space.
 Fock recognized the hydrogen bound solution is a 3-dimensional sphere embedded in a 4-dimensional  space \cite{fock}. He found that the discrete spectrum in {\em P}-space is the Riemann geometry with constant positive curvature, and the continuous spectrum is the Lobachevskii geometry with the constant negative curvature \cite{takhtajan}. For practical applications in {\em P}-space, there is a singularity in the Coulomb kernel of the integral eigenvalue equation which causes numerical trouble \cite{bethe,flugge,bj}. Lande discovered a subtraction method to regularize the singularity \cite{lande}. The Lande method was adopted in several subsequent applications
 \cite{chu,nor,gau}. Recently, we improved the Lande subtraction method by using a reasonable small momentum upper limit instead of mathematical infinity. The accuracy and efficiency are both significantly enhanced.
 We showed
 that the time-dependent Schr\"{o}dinger equation (TDSE) of a hydrogen atom under irradiating of intense laser pulse can be calculated by using the generated {\em P}-space  eigenset of the hydrogen atom. The photoelectron spectra and high-order harmonic generations were obtained without cooperating to other approximations \cite{j1}.

   However, the Lande regularization method has never been straightforward in coding, and application was limited. A simpler formalism than the Lande subtraction method will be useful for  working in momentum space. In this paper, we developed
  a simple but highly accurate method to construct the numerically complete eigenset
  of SAE atom in momentum space.
 The singularity in {\em P}-space Coulomb kernel can be bypassed with numerical quadrature. The SAE eigenset can then be solved easily.
 The accuracies of energy levels and standard deviations of wave functions are both much improved compared with the Lande method. The development may facilitate
 the momentum space applications in quantum mechanical problems.
 Next, in Sec.II,  the formalism was outlined.
 In Sec.III, the results and accuracy of hydrogen and SAE helium were presented.
 In Sec. IV, we showed two calculations of a hydrogen atom under strong-field laser pulse with the eigenset and compared to published results to demonstrate the usage
 of our method.
  This is followed by  conclusions in Sec. V.  Atomic units ( $m_{e}=e=\hbar=1$) are used throughout this paper.
\section{Method of SAE eigenstate calculation}
 In the single-active electron model potential for atoms \cite{sae},  the effective potential contains the nuclear Coulomb potential $-Z/r$, and the short-ranged potential  $V_{short}(r)$  from the screening of other electrons. Let an eigenstate be written as
\[
\Phi_{nlm}(\vec p)=\frac{\chi_{nl}(p)}{p}\,Y_{lm}(\Omega_{p}), \]
then we have \cite{bethe,flugge,bj}
\beq
[ \frac{p^2}{2}-E ] \Phi(\vec p)+\int W(\vec p-\vec q)\Phi(\vec
q) d^3q=0 \,.
\label{saeq}
\eeq
Where we have defined
\bqn
 \Phi(\vec p) &=& 1/{(2\pi)^{\frac{3}{2}}} \int\Psi(\vec
r)e^{-i\vec p \cdot \vec r} d^{3}r \,, \non \\
W(\vec Q) &=& 1/(8\pi^3) \int_{r=0}^{r=R_m}r^{2}\,dr \int d\Omega_{r} [\frac{-Z}{r}+V_{short}(\vec r)] e^{-i\vec Q \cdot \vec r}
\, ,
\eqn
with ${\vec Q}={\vec p}-{\vec q}$.
Instead of $r\in (0,\infty)$ in radial integration mathematically, we numerically integrate it
up to a large enough upper limit $R_m$ where the short-ranged potential
vanished. For the SAE atomic potential, we use
\beq
 V_{short}(r)=-\frac{a_{1}e^{-a_{2}r}+a_{3}re^{-a_{4}r}+a_{5}e^{-a_{6}r}}{r},
 \eeq
with parameters $a_{1}=1.231,a_{2}=0.662,a_{3}=-1.325,a_{4}=1.236,a_{5}=-0.231,a_{6}=0.480$
for helium atom \cite{sae}. The energy levels  were calibrated to NIST database \cite{nist}.
We obtain
\bqn
W(\vec Q)&=& W_{Coul}(\vec Q)+W_{short}(\vec Q), \\
W_{Coul}(\vec Q)&=&\frac{Z}{2\pi^2}\frac{\cos(QR_m)-1}{Q^2} ,\\
W_{short}(\vec Q)&=&\frac{-1}{2\pi^2} \{ \frac{a_{1}}{a_{2}^2+Q^2}+\frac{2a_{3}a_{4}}{(a_{4}^2+Q^2)^2}
+\frac{a_{5}}{a_{6}^2+Q^2} \}.
\eqn
The kernel $W({\vec p}-{\vec q})$ depends only on $p, q$ and the angle $\theta$ between
 ${\vec p}$ and ${\vec q}$. We can make the expansions :
 \bqn
 W_{Coul}({\vec p}-{\vec q})&=&\sum_{l=0} (2l+1)a_{l}(p,q) P_{l}(\xi),\\
 W_{short}({\vec p}-{\vec q})&=&\sum_{l=0} (2l+1)b_{l}(p,q) P_{l}(\xi),
 \eqn
  where $\xi=\cos\theta$, and $P_{l}(\xi)$ is the Legendre polynomial of order $l$. Let
 $z=\frac{p^2+q^2}{2pq}\geq 1$, then
\bqn
a_{l}(p,q)&=&\frac{Z}{8 \pi^2 pq}\int_{-1}^{1} d\xi \frac{P_{l}(\xi)}{z-\xi}[\cos({\sqrt{2pq(z-\xi)}}R_{m})-1], \\
b_{l}(p,q)&=&\frac{1}{2} \int_{-1}^{1} d\xi P_{l}(\xi)W_{short}({\vec p}-{\vec q}).
\eqn
The known Coulomb singularity occurs when ${\vec Q}=0$, that is, $ p=q $ and
$\theta= 0$. In the calculation of Eq.(9), simply apply the quadrature
 of interior grids that excludes the point $\xi =1$ (i.e., $\theta= 0$) will bypass the trouble of the singularity \cite{chiu}. The eigenvalue equation becomes
\beq
(\frac{p^2}{2}-E)\chi_{nl}(p)+4\pi p\int q\,dq[a_{l}(p,q)+b_{l}(p,q) ]\chi_{nl}(q)=0 .
\eeq

In the calculation, the denser grids in small value than in large value will be advantageous
numerically.  A good choice is the nonlinear mapping $p(x)=L\frac{1+x+\beta}{1-x+\alpha}$, where $x\in (-1,+1)$ and $L,\alpha,\beta$ are adjustable parameters that are not sensitive to the results.  We use the Chebyshev  grids, $\{x_{1},x_{2},\cdots,x_{N}\}$, where $\{x_i \}$ are roots of the Chebyshev polynomial $T_{N}(x)$, and weight
$W_t={\pi}/N$. With the  grids $\{x_i\}$ , the quadrature rule is given by
\beq
\int\frac{ F(x)}{\sqrt{1-x^2}}dx\approx W_t\sum_{i=1}^{N} F(x_i).
\eeq

For the hydrogen atom, the eigenvalue equation becomes
\beq
\sum_{j=1}^{N} \left\{ (\frac{p_{i}^2}{2}-E)\delta_{i,j}+4\pi p_ip_j[ a_{l}(p_i,p_j)+b_{l}(p_i,p_j)] p'_{j}W_{t} \right\}\chi_{nl}(p_j)=0,
\eeq
where $p'(x_j)\equiv p'_{j}=\frac{dp(x)}{dx}|_{x=x_j}$.
Let
\beq
 \chi_{nl}(p)= \frac{u_{nl}(x)}{\sqrt{p'(x) }} ,
 \eeq
and expand $u_{nl}(x)$ in terms of the cardinal function $g_{j}(x)$,
\beq
u_{nl}(x)\approx\sum_{j=1}^{N}u_{nl}(x_j)g_{j}(x),
\eeq
where the cardinal function  is \cite{boyd}
\beq
 g_j(x)=\frac{T_{N}(x)}{T'_{N}(x_j)}[x-x_j]^{-1},
\eeq
and $ g_j(x_k)=\delta_{j,k}.$  The eigenvalue equation becomes
  \beq
 \sum_{j=1}^{N} \left\{ (\frac{p_{i}^2}{2}-E)\delta_{i,j}+4\pi p_ip_j\sqrt{p'_{i}p'_{j}}[ a_{l}(p_i,p_j)+b_{l}(p_i,p_j)   ]W_{t}
  \right\} u_{nl}(x_j)=0.
  \label{sae-atom}
 \eeq
The matrix eigenvalue problem is real-symmetric. A solver is available
from LAPACK \cite{lapack}.
\section{Method of the time-dependent Schr\"{o}dinger equation}
 To demonstrate an application of the generated eigenset, we describe the method
 to solve the time-dependent Schr\"{o}dinger equation (TDSE) for a hydrogen
 atom under the interaction of a linearly polarized pulse. The method for
 other SAE atom  is similar.   There are
 several available packages in {\em R}-space for calibration \cite{qprop2,muller}.
 Let the electric field and vector potential be
\bqn
{\vec E}(t)&=&\hat{z}{E_m}\sin^2(\frac{\pi t}{T})\cos(\omega t+\phi)\equiv \hat{z} E(t)\, ,t\in [0,T],\\
{\vec A}(t)&=&-\hat{z}\int_{0}^{t} E(t')dt',
\eqn
where $E_m$ is the peak electric field amplitude, ${E_m}^2$ is the peak intensity, $\phi$ is the carrier-envelope phase, $\omega$ is the laser frequency and $T$ is the pulse duration.  The time-dependent interaction hamiltonian can be written as
\beq
H'(t)=A(t) p_{z} .
\eeq
The split-operator-algorithm for wavepacket time evolution is
 \bqn
 |{\Psi}(t+\Delta)>&=&e^{-iH_{0}\frac{\Delta}{2}}
 e^{-iH'(t')\Delta}e^{-iH_{0}\frac{\Delta}{2}}
 |{\Psi}(t)> + O(\Delta^{3}),\,{\rm with} \\
  t'&=& t + \frac{\Delta}{2}. \non
 \eqn
 The time marching scheme for $t$ to $t+\Delta$ contains the following  steps  :
\subsection{  $|{\Psi}_{1}(t)>\equiv e^{-iH_{0}\frac{\Delta}{2}} |{\Psi}(t)> $}
Let
\beq
{\Psi}(p,\Omega;t)\equiv \sum_{l=0}^{L_m} f_{lm}(p;t)Y_{lm}(\Omega)/p,
\eeq
where the initial magnetic quantum number $m$ is conserved under linearly polarized pulse, and
angular momenta from $l=0$ to $l=L_{m}$ being used,
\bqn
|{\Psi}_{1}(t)>&\equiv&e^{-iH_{0}\frac{\Delta}{2}}| {\Psi}(t)>,\non \\
 &=&\sum_{n,l}e^{-iH_{0}\frac{\Delta}{2}}|nlm><nlm| {\Psi}(t)>,\non \\
 &=&\sum_{n,l}e^{-iE_{nl}\frac{\Delta}{2}}|nlm>\int qdqd\Omega \chi_{nl}^{*}(q)Y_{lm}^{*}(\Omega) {\Psi}(q,\Omega;t),\,{\rm with}\non \\
{\Psi}_{1}(p,\Omega;t)&\equiv&\sum_{l=0}^{L_m}f^{(1)}_{lm}(p;t)Y_{lm}(\Omega)/p , \,{\rm then}\\
 f^{(1)}_{lm}(p;t)&=&\sum_{n}e^{-iE_{nl}\frac{\Delta}{2}}\chi_{nl}(p)\int dq\chi_{nl}^{*}(q)f_{lm}(q;t).
 \eqn
 Using the Gauss-Chebyshev quadrature, we have
 \bqn
f^{(1)}_{lm}(p_{j};t)&=&\sum_{k}A^{(l)}_{j,k}f_{lm}(p_{k};t), \non \\
A^{(l)}_{j,k}&=&\sum_{n}e^{-iE_{nl}\frac{\Delta}{2}}\chi_{nl}(p_j)\chi_{nl}^{*}(p_k) p'_{k}\,W_{t}.
\eqn
The matrix $A^{(l)}$  for each $(l,m)$ is time-independent and is m-independent, $l \in [0,L_{m}]$ with the maximum angular momentum $L_m$. It can be constructed outside of the time marching loop.
\subsection{ $|\Psi_{2}(t)>\equiv e^{-iA(t+\frac{\Delta}{2}) p\cos\vartheta \Delta}
|{\Psi}_{1}(t)>$}
 Let's write
\beq
\Psi_{2}(p,\Omega;t)\equiv \sum_{l=0}^{L_m} f^{(2)}_{lm}(p;t)Y_{lm}(\Omega)/p.
\eeq
By using the Rayleigh's formula
\beq
e^{-iA(t') p\cos\vartheta \Delta} =\sum_{k=0}(-i)^{k}(2k+1) j_{k}[A(t') p \Delta ]P_{k}(\cos\vartheta),
\eeq
where $t'=t+\frac{\Delta}{2}, j_{k}$ is the spherical Bessel function and $P_{k}$ is the Legendre polynomial; especially for the case of $m=0$, we can derive
\beq
f^{(2)}_{l0}(p;t)=\sum_{l_{1},l_{2}}(-i)^{l_1}j_{l_1}[A(t') p \Delta](2l_{1}+1)\sqrt{(2l+1)(2l_{2}+1)}f^{(1)}_{l_{2}0}(p;t)\left( \begin{array}{ccc} l& l_{1} & l_{2} \\  0& 0&  0  \end{array} \right)^{2},
\eeq
where the last symbol in the above equation is square of a Wigner-3j symbol.
\subsection{$|{\Psi}(t+\Delta)> \equiv e^{-iH_{0}\frac{\Delta}{2}} |{\Psi_2}(t)> $  }
 This step repeats the algorithm of step {\bf A}.
\subsection{Above-threshold-ionization and photoelectron angular distribution}
  At the end of pulse, we write the final wave function as
\beq
 \Psi(t=T)\equiv\Psi_{f}(\vec p)=\sum_{l}f_{lm}(p;T)Y_{lm}(\Omega)/p .
\eeq
For the above-threshold-ionization (ATI) and photoelectron angular distribution
(PAD), we need the wave packet $\Psi^{conti}(\vec p)$ in the continuum,
\bqn
\Psi^{conti}(\vec p)&=&\Psi_{f}(\vec p)-\sum_{nlm}\Psi^{bound }_{nlm}(\vec p) \int\Psi^{bound}_{nlm}(\vec q)^{*}\Psi_{f}(\vec q)d^3q, \\
& \equiv & \sum_{l}g_{lm}(p)Y_{lm}(\Omega)/p , \\
g_{lm}(p)&=&f_{lm}(p)-\sum_{n}\chi^{bound}_{nl}(p)\int dq \chi^{bound}_{nl}(q)^{*}f_{lm}(q),
 \eqn
where $\chi^{bound}_{nl}(p)$ are radial wave function of bound state with quantum number $(nl)$.
Let the ionization probability be $\cal P$, then
\bqn
{\cal P}&\equiv&\int_{\epsilon=0}^{\infty} \frac{\partial \cal P}{\partial\epsilon}d{\epsilon}=\int  |\Psi^{conti}(\vec{p})|^2 d^{3}p ,\, {\rm where}\non \\
\frac{\partial \cal P}{\partial\epsilon}&=& \int\int p|\Psi^{conti}(\vec{p})|^2 d\Omega \non\\
&=&\sum_{l} |g_{lm}(p)|^2/p.
\eqn
The photoelectron angular distribution (PAD) is conveniently expressed in 2-dimensional momentum plot.
Let $p_{\|}$ and $p_{\bot}$ be the component of momentum $\vec p$ in parallel and in perpendicular to the polarization axis, $\vartheta$ be the angle between  $\vec p$ and the polarization axis, then
$p_{\|}=p\cos{\vartheta}$, and $p_{\bot}=\sqrt{p^2_{x}+p^2_{y}}=p\sin{\vartheta}$. The ionization probability $\cal P$ will be expressed as
\bqn
{\cal P}&\equiv&\int \frac{\partial^2\cal P}{ \partial p_{\|} \partial p_{\bot} } dp_{\|}dp_{\bot},\, {\rm where} \non\\
 \frac{\partial^2\cal P}{ \partial p_{\|} \partial p_{\bot} }&=&\int p_{\bot}\cdot |\Psi^{conti}(\vec{p},t=\infty)|^2 d\varphi.
\eqn

\section{Results and discussions}
We apply the present momentum space method to the hydrogen atom first.
The results can be compared to the exact solutions.
Some low-lying hydrogen radial wave functions $F_{nl}(p)\equiv\chi_{nl}(p)/p$ were listed in Ref.\cite{bj}. Those bound state $F_{nl}(p)$ which are not listed can be derived by using the Gegenbauer polynomials.  The exact energy level $E(nl)_{exact}$ of eigenstate $\Phi_{nlm}(\vec p)$ is equal to $-1/(2n^2)$.
We showed the error for calculated energy level $\Delta E=|E(nl)-E(nl)_{exact}|$
 to demonstrate the accuracy.
In Table I, we listed the accuracy of some calculated hydrogen low-lying energy levels.
The tabulated results are of 512 and 1024 grid numbers with momentum maximum $p_{max}=50$.
For the N=512 case, the relative errors are in order of $10^{-10}\sim 10^{-4}$. The worst one is the ground state with error $3.46\times 10^{-4}$ which is correct up to the fourth place after the decimal point. For $N=1024$, it is accurate up to the sixth decimal place.
Because the ground state wave function $F_{1s}(p)=\sqrt{\frac{32}{\pi}}\frac{1}{(1+p^{2})^2}$
decays slowly with $p$, a larger $p_{max}$ can give higher accuracy. Other excited state wave function is located in a small region of $p$ such that $p_{max}=50$ gave good accuracy.
Doubling the grid number, the accuracy improves in nearly two orders of magnitude.
Compared with $N=2048$, $p_{max}=100$ of case I, Table I in Ref.\cite{j1} of our previous complicated Lande subtraction method, the present accuracy with half grid number has been competitive
or even better. The results show that high accuracy  in $E(nl)$ was reached at a modest number of grids.

 For the accuracy of calculated wave functions, we use the root-mean-square (rms) deviation for $ \chi_{nl}(p)$,
\beq
 \Delta \chi_{nl}=\sqrt{\frac{1}{N}\int dp [\chi_{nl}(p)-\chi_{nl}(p)_{exact} ]^2},
 \eeq
 where $\chi_{nl}(p)_{exact}$ are obtained through the analytical wave functions \cite{bj}.
 In Table II, we listed the rms deviations of the first four low-lying states for $L=0,1,2,3$.
The high accurate results were shown. The rms deviation is in the order of $10^{-10}\sim 10^{-6}$ with grid number N=1024. Accuracy improves about two orders by increase N=512 to 1024. Again, the accuracy is competitive or even better than the Lande method. The rms deviation represents typical deviation of wave function at each grid point. This accuracy is good
 enough for the further applications such as time-dependent calculations. In Table III, we tabulated  errors of the first four low-lying energy levels of $L=0,1,2,3$ for helium with SAE potential. We use the results generated by the high accurate eigenstate solver \cite{cpc} for comparison. Compare with the Case I, Table II of Ref.\cite{j1} at
$N=1024$ and $p_{max}=50$, we can find more than two orders of improved accuracy were reached. We can claim that the excellent accuracy has been reached.

 In the first TDSE check, we calculate the ground state hydrogen atom
 under the interaction of a 800nm, FWHM $10 fs$ Sine-square envelope pulse in
 the electric field with  peak intensity $1.0\times 10^{14} W/cm^2$.
 The corresponding vector potential is calculated through $A(t)= -\int_{-\frac{T}{2}}^{t}E(t')dt'$.
  Fig.1a is the above-threshold-ionization (ATI) spectrum which is to compare with the Fig.10a of \cite{cdl}. The shape and
 magnitude are totally agreed with each other. Fig.1b is the corresponding photoelectron angular distribution (PAD). PAD shows the density distribution
 for photoelectron with momentum in parallel and in perpendicular to the polarization direction. The depicted photoelectron kinetic energy of $0\sim7\, eV$ in Fig.1a corresponds to
 the electron momentum in $0\sim0.72\, a.u.$ The first 3 peaks group in ATI corresponds to the fan structure  for $0 < p < 0.4$. The next 3 ATI peaks are rings in $0.4 < p < 0.61$.
 The number of zero stripes in the PAD  shows the corresponding angular momentum. For the  first fan group, there are 3 zero stripes in PAD which the corresponding Legendre polynomial is of order 3, that is, the photoelectron angular momentum $l= 3$. In this calculation, we use $l\in[0,31]$ of the generated eigenset with $p_{max}= 50$ at $1024$ grids. The convergence is checked with $L_{max}$ increased up to $47$.
 In the next case, we calculate hydrogen ground state under 535nm, 20-cycle, peak intensity $2\times 10^{13} W/cm^2$, Sine-squared envelope vector potential pulse. This is to ccompare with Fig.2 of \cite{qprop2}. In our Fig.2a, we showed the ATI. We can see the results agree well with \cite{qprop2}. Also showed is the corresponding PAD in Fig.2b. The PAD contains information of photoelectron detailed distribution in momentum space.

\section{Conclusions}
 By using the simple and straightforward quadrature and nonlinear mapping in momentum grids, the accuracy of SAE low-lying states in momentum space
 has been competitive or better than our previous complicated Lande method \cite{j1}.
 In which, we improved the Lande subtraction by analytic extension of momentum
range from a finite upper limit $p_{max}$ to infinity.  Much more complicated formalism is required than current method.
  For the ground state, its momentum space wave function vanishes slowly with $p$. Hence an increase of $p_{max}$  is helpful for better accuracy.
For the excited states and continuous states, the wave functions squeezed into the small momentum
region. A modest $p_{max}$ and the denser grids near origin than in the large
value of momentum  are necessary for the accurate spectrum.
This requirement was fulfilled by the nonlinear mapping.
Changing the mapping parameter in $p=p(x)$ can adjust the grids distribution and improve the accuracy a little but not in a sensitive way.  Increasing the number of grid points will improve the accuracy a lot by comparison the results of $N=512$ and $N=1024$.
High precision is capable with a large $p_{max}$ and increase of grid points. The data in  Tables I and III showed that the errors in  energy levels were quite small indeed. The rms deviation of wave function showed in Table II elucidated the good accuracy was reached.
 The calculation in this paper was carried out in a desktop personal computer with
 Intel Xeon(R) E5-2630v3 cpu. Results of Table I and II with $N=512$ cost wall time of $\sim 29\,$ second, and the cpu time scales to $\sim N^2$.
Application of the present momentum space method in the study of strong light generated photoelectron has the advantage over coordinate space method \cite{chu,jcp,j1}. We showed two TDSE examples of a hydrogen atom under laser pulses. The agreements with published results are very good. This is just to demonstrate the applicability of {\em P}-space method because of calculations in this category  were not general.
We have developed the TDSE code in GPU. The capability of solving TDSE with
circular polarized pulse has been realized.
Theoretical investigation on recently interested problems such as mid-infrared pulse on atoms, low-energy structure photoelectron spectra \cite{les},
dichroism \cite{dicho}, spin polarization of photoelectron \cite{spin} etc. are under progress.

{\bf Acknowledgment}
    The work is supported by the Ministry of Science and Technology, Taiwan under contract numbers
    MOST105-2112-M-009-003 and MOST105-2811-M-009-060.
\newpage
\begin{table}[ht]
\caption{The magnitude of errors $\Delta E(i),i=1,2,3,4$
for the hydrogen lowest four levels of $L=0,1,2,3$.
$\Delta E$  is defined as $|E(nl)-E(nl)_{exact}|$, with
$E(nl)_{exact}=-1/ (2n^2)$ and $E(nl)$ denotes the calculated energy level.
The numbers of  grid points  are 512,1024 respectively,
and $p_{max}=50$. 3.46E-4 designates $3.46\times 10^{-4}$. }
{\begin{tabular}{c|cccccccc}  \toprule
 state and & \multicolumn{2}{c}{ $\Delta E(1)$}
 &\multicolumn{2}{c}{$\Delta E(2)$}
 &\multicolumn{2}{c}{$\Delta E(3)$}
 &\multicolumn{2}{c}{$\Delta E(4)$} \\
\cline{2-3} \cline{4-5} \cline{6-7} \cline{8-9} {grid points}
&512&1024&512&1024&512&1024&512&1024 \\
  \hline
$L=0$ &3.46E-4& 3.77E-6& 5.74E-5& 3.93E-7& 1.80E-5& 1.12E-7& 7.75E-6&4.65E-8\\
$L=1$ &3.57E-6& 1.41E-8& 1.34E-6& 5.04E-9& 6.12E-7& 2.25E-9& 3.24E-7&1.18E-9\\
$L=2$ &2.36E-8& 1.8E-11&1.46E-8& 1.0E-11&8.67E-9& 6.3E-12&5.41E-9&4.2E-12\\
$L=3$ &1.1E-10&1.5E-12&9.5E-11&6.0E-13&7.0E-11&5.4E-13&2.6E-10&2.1E-10 \\
 \botrule
 \end{tabular}
 \label{ta1}}
\end{table}
\begin{table}[ht]
\caption{The  root-mean-square deviations $\Delta \chi_{nl}(i),i=1,2,3,4$
 of wave function for the hydrogen lowest four states of $L=0,1,2,3$.
The numbers of  grid points  are 512,1024 respectively,
and $p_{max}=50$. }
{\begin{tabular}{c|cccccccc} \toprule
 state and & \multicolumn{2}{c}{ $\Delta \chi(1)$}
 &\multicolumn{2}{c}{$\Delta \chi(2)$}
 &\multicolumn{2}{c}{$\Delta \chi(3)$}
 &\multicolumn{2}{c}{$\Delta \chi(4)$} \\
\cline{2-3} \cline{4-5} \cline{6-7} \cline{8-9} {grid points}
&512&1024&512&1024&512&1024&512&1024 \\
  \hline
 $L=0$ & 4.49E-5& 6.69E-7&4.05E-5&3.85E-7& 4.02E-5&3.35E-7 & 4.03E-5&3.26E-7 \\
 $L=1$ & 4.92E-6& 3.28E-8&4.39E-6&2.40E-8& 4.24E-6&2.09E-8 &4.23E-6 &1.98E-8 \\
 $L=2$ & 3.18E-7& 2.62E-9&2.67E-7&3.06E-9& 2.24E-7&3.33E-9 &1.98E-7 &3.57E-9 \\
 $L=3$ & 4.51E-8& 5.2E-11&5.73E-8&6.4E-11 &6.54E-8 &7.1E-11 &7.39E-8 &9.4E-10 \\
  \botrule
 \end{tabular}
 \label{ta2} }
\end{table}
\begin{table}[ht]
\caption{The magnitude of errors  $\Delta E(i),i=1,2,3,4$  for the SAE Helium
lowest four levels of $L=0,1,2,3$.   $\Delta E$
 is defined as $|E(nl)-E(nl)_{R}| $
where $E(nl)_{R}$ is the level accurately calculated by the solver of Ref.\cite{cpc}
and  $E(nl)$ denotes the calculated energy level by current momentum space method.
The numbers of  grid points  are 512,1024 respectively, and $p_{max}=100$. }
{\begin{tabular}{c|cccccccc} \toprule
 state and & \multicolumn{2}{c}{ $\Delta E(1)$} &\multicolumn{2}{c}{$\Delta E(2)$} &\multicolumn{2}{c}{$\Delta E(3)$}&\multicolumn{2}{c}{$\Delta E(4)$} \\
\cline{2-3} \cline{4-5} \cline{6-7} \cline{8-9} {grid points}
&512&1024&512&1024&512&1024&512&1024 \\
  \hline
$L=0$ &1.54E-3& 8.40E-5& 1.70E-4& 6.92E-6& 4.66E-5& 1.81E-6& 1.89E-5& 7.22E-7\\
$L=1$ &9.30E-6& 1.03E-7& 3.37E-6& 3.59E-8& 1.52E-6& 1.57E-8& 8.02E-7& 8.16E-9\\
$L=2$ &4.44E-8& 3.7E-10&2.69E-8& 9.0E-11&1.59E-8& 4.2E-11&9.92E-9& 2.0E-11\\
$L=3$ &4.4E-10&2.5E-10&3.0E-10&1.3E-10&1.9E-10&7.1E-11&1.3E-10&4.2E-11 \\
 \botrule
 \end{tabular}
 \label{ta3} }
\end{table}
\newpage
\begin{figure}[ph]
\centering
\mbox{\rotatebox{-90}{\myscalebox{
\includegraphics{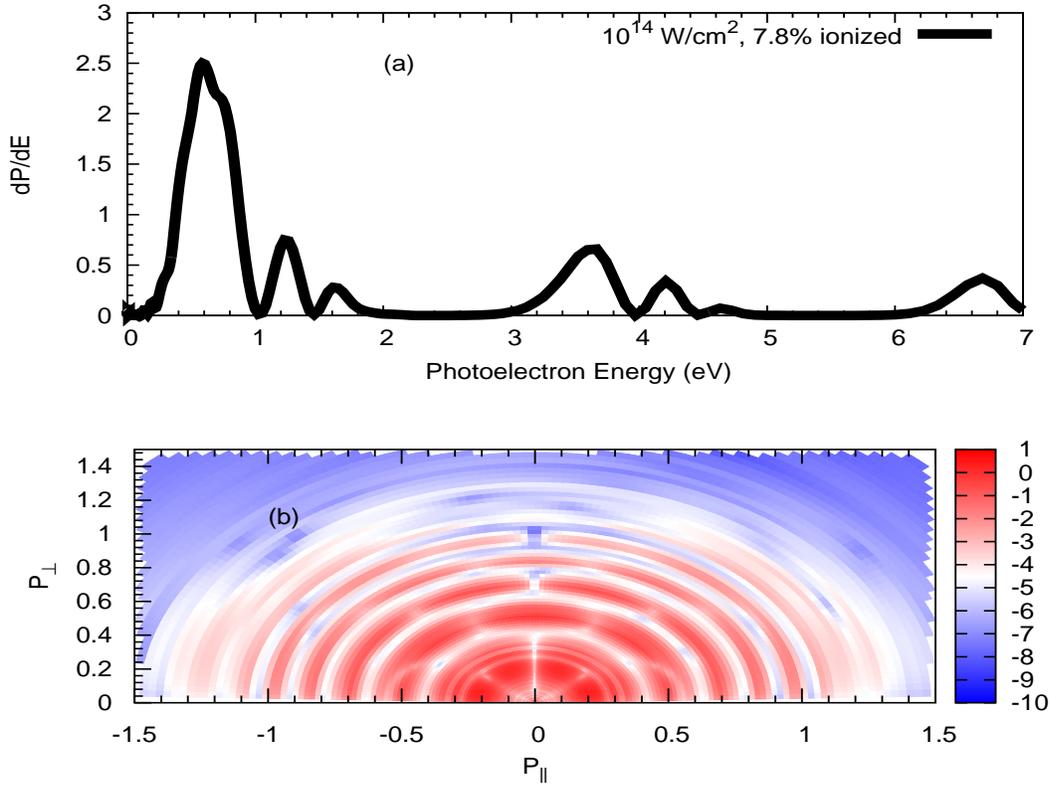}} }}
\caption{ (Color online) (a) The ATI spectrum of the hydrogen atom
 under the interaction of a 800nm, FWHM $10 fs$ Sine-square envelope pulse in
 the electric field with  peak intensity $1.0\times 10^{14} W/cm^2$. The plot
 is to compare with Fig.10a of Ref.\cite{cdl}.
 (b) The corresponding PAD.
}
\end{figure}
\begin{figure}[ph]
\centering
\mbox{\rotatebox{-90}{\myscalebox{
\includegraphics{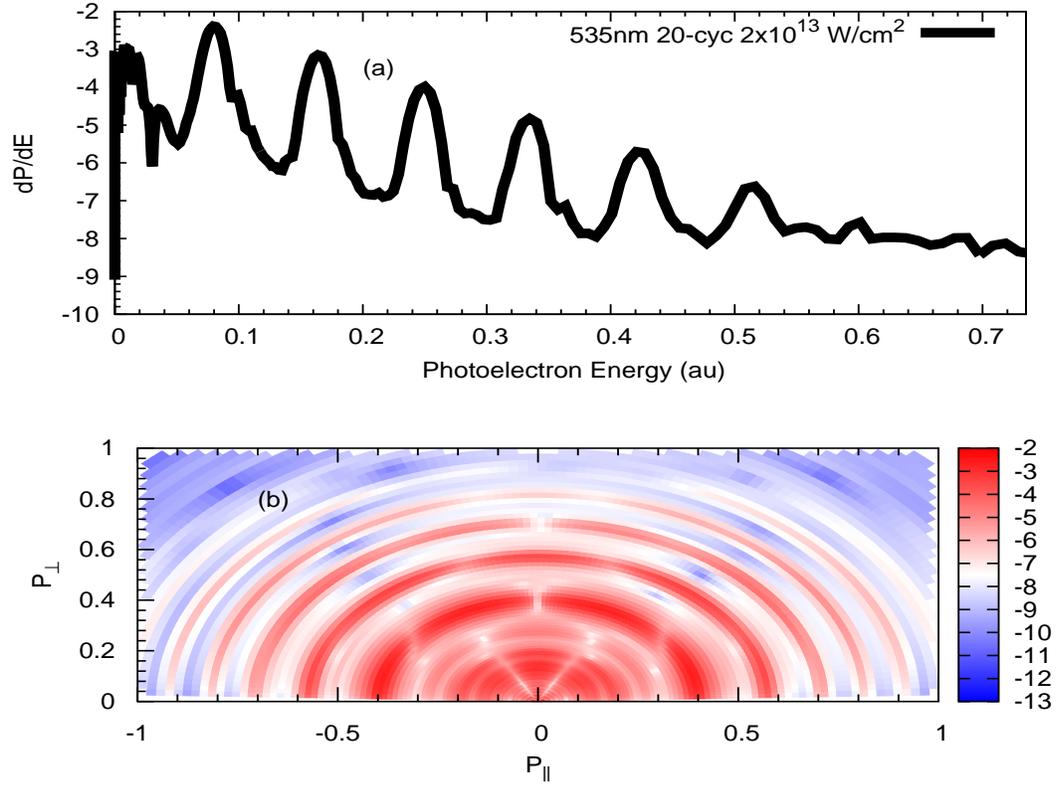}}}}
\caption{ (Color online) (a) The hydrogen atom under 535nm, 20-cycle, peak intensity $2\times 10^{13} W/cm^2$, Sine-squared envelope pulse in vector potential. This is to compare with Fig.2 of \cite{qprop2}. The vertical axis is in exponential scale.
(b) The corresponding PAD. }
\end{figure}
\end{document}